\documentclass[aps,prl,showpacs,floats,10pt,twocolumn]{revtex4}
\usepackage{graphicx}
\usepackage{amsfonts}
\usepackage{times}
\begin{document}
\def\E{{\bf E}}

\def\T {|q\rangle \langle q|} \def\dip{{3 |{\bf r}\rangle \langle {\bf
      r}| -1 \over 4 \pi r^3}}

\def\P{{\bf P}}

\def\D{{\bf D}}

\def\q{{\bf q}}

\def\rr{{\bf r}}

\def\dV{{\; \rm d^3}{\bf r}}

\def\div{{{\nabla \cdot} }}

\def\curl{{{\rm curl}\; }}

\def\p{{\bf p}}

\def\rhat{\hat {\bf r}}

\def \dv{\; d^3\rr}

\title{Monte Carlo Simulation of a Model of Water}

\author{A. C. Maggs} \affiliation{ Laboratoire de Physico-Chime
  Th\'eorique, UMR CNRS-ESPCI 7083, 10 rue Vauquelin, 75231 Paris
  Cedex 05, France.  }
\begin{abstract}
  We simulate TIP3P water using a constrained Monte Carlo algorithm to
  generate electrostatic interactions eliminating the need to sum over
  long ranged Coulomb interactions. We study discretization errors
  when interpolating charges using splines and Gaussians.  We compare
  our implementation to molecular dynamics and Brownian dynamics
  codes.
\end{abstract}
 \maketitle

The TIP3P model of water \cite{klein} is often used to study the
accuracy of algorithms for atomistic simulation. The model has a
single Lennard-Jones center representing an oxygen atom together with
three charges $(-0.834, +0.417, +0.417)$ arranged in a triangle.  The
oxygen-hydrogen bond length is 0.9572\AA\, the angle between bonds is
$104.52^\circ$.  Accurate simulation of this model is surprisingly
challenging: The bare electrostatic interaction between oxygen atoms
at a separation of 2.75\AA, is over $100\, k_BT$.  Small errors in the
representation of the electrostatic potential lead to significant
errors in the total energy due to large cancellations; the binding
energy per hydrogen bond is only $7\, k_BT$.

Many molecular dynamics codes for the simulation of large numbers of
charges are based on Poisson solvers. The codes interpolate charges to
a cubic grid and then calculate the electrostatic energy via fast
Fourier transform \cite{darden} or multigrid \cite{sagui,shaw}.  The
principle difficulty is controlling errors in the Coulomb interaction
using high order interpolation. One requires a relative error of at
most $\sim 10^{-4}$. In this article we present a Monte Carlo
algorithm for simulation at this level of accuracy. We avoid solving
the Poisson equation by generalizing an algorithm 
which generates the Coulomb interaction between particles using Monte
Carlo evolution of the electric field.  Previous codes using this
local algorithm have been of low accuracy, sufficient for the study of
lattice gasses \cite{duncan} or charges interacting through an
implicit solvent \cite{igor}. They were still far from the accuracy
needed for the simulation of TIP3P.  This articles considers the
modifications necessary to the algorithm in order to reliably simulate
standard atomistic models.

There were three important sources of error in the energy functions
used in previous work with local electrostatics algorithms:
\begin{itemize}
\item use of low order interpolation leading to distorted charge
  distributions
\item aliasing errors in the lattice Green functions leading to a
  self-energy with the periodicity of the lattice
\item low order discretization of the lattice Green function leading
  to anisotropy in the effective interactions.
\end{itemize}

In many codes interpolation of charges from the continuum to the cubic
grid is performed with splines. A one-dimension $n$-spine is a set of
$n$ polynomials of order $n-1$.  These polynomials give the quantity
of charge which is deposited on $n$ consecutive sites of the lattice
as a function of the position of the particle, $f_i(x)$, $1\le i \le
n$.  Linear interpolation corresponds to a 2-spline. In three
dimensions one takes the product of  splines in the $x$,
$y$ and $z$ directions, interpolating a charge to $n^3$ lattice site,
thus $f_{\bf l}(\rr)= f_i(x) f_j(y) f_k(z)$ for $\rr=(x,y,z)$ and
${\bf l}=(i,j,k)$. Splines have several useful properties for
interpolation: They conserve total charge exactly; they are smooth
with $n-2$ continuous derivatives.  With Fourier solvers splines work
well if one takes $n \ge 4$ \cite{darden}.

An alternative to splines is interpolation with truncated Gaussians
\cite{darden,shaw}.  Consider interpolating a unit charge to a
one-dimensional grid with Gaussian interpolation:
\begin{math}
  f_i(x)= \exp{(- (x-i)^2/{2 \sigma^2})}/\sqrt{2 \pi }\sigma
\end{math}. The total interpolated charge can be evaluated for $\sigma$ large
with the Poisson resummation formula:
\begin{math}
  q_{int}= \sum_i f_i(x) = \sum_{p} \tilde f(2 \pi p)
\end{math}
where $\tilde f$ is the (continuous) Fourier transform of $f_i(x)$.
We find
\begin{math}
  q_{int} \sim 1+ 2\cos(2 \pi x) e^{-2 \pi^2 \sigma^2 } \label{error}
\end{math}.
Already for $\sigma=1$ errors in charge conservation are $O(10^{-8})$.
In practice one truncates beyond $\lambda \sigma$ where $\lambda \sim
4-5$, leading to an additional error which varies as
$e^{-\lambda^2/2}$.

In order to study the various errors generated with lattice Monte
Carlo algorithms for the electrostatic energy consider the interaction
between two particles placed at $\rr$ and $\rr'$.
\begin{eqnarray}
  U(\rr,\rr') &=& \sum_{{\bf l},{\bf m}} f_{\bf l}(\rr) G({\bf l}-{\bf m})
  f_{\bf m}(\rr') 
 \label{U0} \\
  &=& \sum_{\bf p} \int {d^3{\bf q} \over (2 \pi)^3} \;  \tilde f(\q-2 \pi {\bf p} ) 
  \tilde f(\q) G(\q)  \nonumber \\
 &\, & \times e^{i \q \cdot (\rr-\rr')+2 \pi i{\bf p}\cdot \rr} 
\label{U1}
\end{eqnarray}
$G({\bf l})$ is the lattice Green function of the interaction between
two sites and $G(\q)$, its Fourier transform, has the periodicity of
the Brillouin zone. $2\pi{\bf p}$ is a vector of the reciprocal lattice.
The integral is over all Fourier space. The particles also have a self
energy  $U(\rr,\rr)/2$.

Consider the contribution ${\bf p}=0$ in eq.~(\ref{U1})
\begin{equation}
  U_{0}(\rr,0) = \int G({\bf q}) \tilde f^2({\bf q})
  e^{i \q \cdot \rr} {\dv \over (2 \pi)^3} \label{Ua} \\
\end{equation}
If $\tilde f$ is Gaussian and $G(r)=1/4 \pi r$, $G(q)=1 /q^2$ so that
\begin{equation}
  U_0 (\rr,0) + {{\rm  erfc}(r/2\sigma) \over 4 \pi r}= \label{erfc}
  {1\over 4 \pi r} 
\end{equation}
This is the central formula for so called ``particle mesh Ewald''
methods \cite{darden,shaw}.  One neglects contributions with ${\bf p}
\ne 0$ and calculates the Coulomb energy as a lattice energy,
eq.~(\ref{U0}), plus a short range correction.  Deviations from
eq.~(\ref{erfc}) occur if the structure factors are not Gaussian: For
splines $\tilde f=\prod_\alpha {\rm sinc}^n(q_\alpha/2)$ with
$\alpha=(x,y,z)$ which has the cumulant expansion
\begin{math}
  \, \tilde f \sim \prod \exp({-n q_\alpha^2/ 24} - {n q_\alpha^4/
    2880 })
\end{math}. Splines converge for large $n$
to Gaussians of width $\sigma^2=n/12$. However interpolation with
splines generates an extra contribution at $q=0$ in eq.(\ref{Ua}) due
to the term in $q_\alpha^4$ in the cumulant.  This leads in real space
to an error which decays as $1/r^5$. The amplitude of this error
decays rather slowly with $n$.

The aliasing error comes from the contributions ${\bf p} \ne 0$.
Consider, for instance the self energy $U(\rr,\rr)/2$ and the
contribution to eq.~(\ref{U1}) from ${\bf p}_1=(1,0,0)$.  Since $\tilde f({\bf
  q})$ decays rapidly in Fourier space the product $\tilde f(\q-2 \pi {\bf
  p}_1) \tilde f(\q)$ is maximum on the boundary of the first Brillouin zone
near $ {\bf q}= \pi (1,0,0)$. If we sum over all symmetry related
lattice reciprocal vectors we find a periodic one body potential
\begin{equation}
V_1\sim \tilde f^2(\pi {\bf p}_1) G(\pi {\bf p}_1) \sum_\alpha \cos {2  \pi r_\alpha}
 \label{sinusoid}
\end{equation} 
Higher order corrections to $V_1$ come from larger ${\bf p}$.  We
compare spline and Gaussian interpolation: For a Gaussian $\tilde
f^2(\pi{\bf p}_1)= e^{- \pi^2 \sigma^2}$, whereas for a $n$-spline we
find $\tilde f^2 = (2/\pi)^{2n} \sim e^{-0.9 n}$.  Requiring $\tilde
f^2 \sim 10^{-4}$ implies that $\sigma \sim 1$ or $n\sim 10$. An
implementation using low order splines with only $n=3$ showed strong
aliasing artefacts \cite{continuum}.  The sinusoidal form of
eq.~(\ref{sinusoid}) permits simple analytic subtraction, but we will
not pursue this point here.

We now turn to errors in the lattice Green function, $G(\rr)$.
Coulomb's law in electromagnetism results from the imposition of a
linear constraint, Gauss' law: $\div \E =\rho$, on a quadratic energy
functional: $U=1/2 \int \E^2 \dv$.  Previous codes that discretized
these equations led to the standard 7-point discretization of the
Laplacian operator:
\begin{equation}
  G^{-1}({\bf q}) = 2\sum_{\alpha=1}^3 (1-  \cos q_\alpha )  \label{disp0}
\end{equation}
Expanding we find
\begin{equation}
 G({\bf q}) = {1\over q^2} + {1\over q^4}\sum_{\alpha=1}^3 {q_\alpha^4\over 12} + \dots
\end{equation}
The presence of terms which involve $q^4_\alpha/q^4$ imply a
correction to $G(\rr)$ which decays as only $1/r^3$.  We now construct
a discretization which converges faster. Consider an energy which is a
quadratic function of $P$ electric field variables $E_i$ where the
subscript includes both positional and directional information.
\begin{eqnarray}
  U_E&=& {1\over 2} \sum_{i,j}^P E_i K_{ij} E_j \label{U2} \\
  &\equiv& EKE/2\nonumber
\end{eqnarray}
We continue with an operator notation for compactness.  We submit this
energy to $L<P$ linear constraints, $c_l$
\begin{equation}
  c_l \equiv \sum_{p=1}^P D_{lp} E_p -e_l=0,\quad  \forall l \label{constraint}
\end{equation}
where $D_{lp}$ is a discretization of the divergence operator at $l$
and $e_l$ is the charge. Stationary points are found by considering
the functional
\begin{math}
  A= U_E - \sum_l \phi_l c_l
\end{math}.  $D$ is a linear operator, we define the adjoint $D^*$.
The variational equations of $E$ are
\begin{equation}
  KE - D^*  \phi=0 \label{variation}
\end{equation}
Solving for $E$ in eq.~(\ref{variation}) and substituting in the
constraint equation eq.(\ref{constraint}) we find a generalized
Poisson equation for the Lagrange multipliers $\phi$
\begin{equation}
  D K^{-1} D^* \phi - e=0
\end{equation}
From this Poisson equation we find that the minimum of the constrained
energy is
\begin{equation}
  U_c = e \phi/2=eG_Ke/2
\end{equation}
with the Green function $G_K^{-1} =D K^{-1}D^*$. We make contact with
electrostatics if we recognize that $D={\rm div}$ implies $D^{*} =
-{\rm grad}$, and $G^{-1}= -\nabla^2$.

We will now generalize these results to non-zero temperatures and show
that the effective interaction between particles is still described by
the Green function $G_K$. The constraints are now imposed by
delta-functions in a partition function
\begin{equation}
  Z= \int \prod_{p=1}^P dE_p e^{-\beta U_E} \prod_{l=1}^L \delta(c_l)
\end{equation}
We decompose the field $E$ into generalized ``longitudinal'' and
``transverse'' components by writing $E= K^{-1} D^{*} \phi + E_t$ and
change integration variables from $E$ to $E_t$. The partition function
then factorizes
\begin{eqnarray}
  Z &=& e^{-\beta U_c}  \int \prod_{p=1}^P dE_{t,p}
 e^{-\beta E_t K E_t/2}\prod_{l=1}^L \delta( DE_t )\nonumber \\
  &=& Z_{K} \times {\rm constant}
\end{eqnarray}
Where $Z_{K}$ is a partition function for particles interacting with
the Green function $G_K$.

Previous implementations \cite{vincent} took the following forms for
the operators $D$ and $K$: $DE$ was the flux out of the site $l$ to
the six nearest neighbor sites. $K$ was diagonal, $K=\delta_{ij}$,
leading to eq.~(\ref{disp0}).  Here we keep the same form for the
operator $D$ but for $K$ we include interactions between neighboring
links on the lattice; for $x$-oriented bonds of the lattice the energy
function is
\begin{equation}
  U_E = {1\over 2} \sum 
  \left \{
    {5\over 6} E^2_{i,j,k|x} + 
    {1\over 12} E_{i,j,k|x}(E_{i+1,j,k|x} + E_{i-1,j,k|x})
  \right \}
  \label{locale}
\end{equation}
with similar expression for the links in the $y$ and $z$ directions.
In Fourier space we find that
\begin{eqnarray}
K(\q) &=& {1\over 6} {\rm diag} ( 5+ \cos q_x, 5+ \cos q_y, 5+ \cos q_z )\nonumber \\
D(\q) &=& (1- e^{i q_x}, 1- e^{i q_y}, 1-e^{i q_z} ) \cdot \nonumber \\
D^*(\q) &=& (1- e^{-i q_x}, 1- e^{-i q_y}, 1-e^{-i q_z} )^T \nonumber 
\end{eqnarray}
where ``${\rm diag}$'' denotes a matrix with the indicated diagonal
elements, so that
\begin{eqnarray}
  G^{-1} 
  &=& D(\q)K^{-1}(\q)D^*(\q)= 12 \sum_\alpha {1-  \cos q_\alpha  \over 5 +  \cos q_\alpha  } \nonumber
  \\
  G(\q)&=& {1\over q^2} + {1\over q^4 }\sum_\alpha {q_\alpha^6\over 240 } + \dots \label{disp2}
\end{eqnarray}
This form of $G$ leads to reduced artefacts in the lattice Green
function; errors now decay as $1/r^5$.

\begin{figure}[htb]
  \includegraphics[scale=.45,] {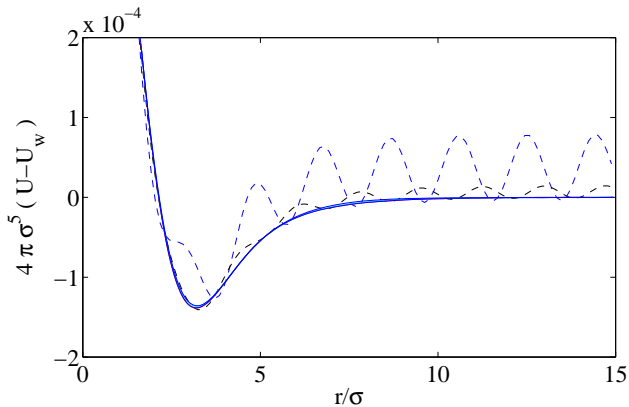}
  \caption{Scaled error in the pair potential  using the energy
    eq.~(\ref{disp2}) for $\sigma=$ 0.90, 1.0, 1.2, 1.4, 1.6.  One
    particle placed at $(0,0,0)$ the second displaced in the direction
    $(1,1,1)$.  Solid lines: Curves for $\sigma=$ 1.2, 1.4, 1.6
    superpose: errors in $G$ dominate. Also included a single curve
    for $\sigma=1.2$ starting at $(0,0,0.5)$ which scales in an
    identical manner.  Dashed lines: $\sigma=$ 0.90, 1.0.
    Oscillations, eq.~(\ref{sinusoid}), from aliasing are also
    important and violate the scaling in $\sigma^5$.}
  \label{fig:v5}
\end{figure}

To calibrate the effective interaction generated by our constrained
algorithm we numerically inverted the Green function
eq.~(\ref{disp2}). We take two interpolated unit charges and measured
the potential between them, Figure~(\ref{fig:v5}), as a function of
$\sigma$ and compared with the (exact) Ewald energy, $U_w$.  We find
collapse of the error when we plot $4 \pi (U-U_w) \sigma^5$ as a
function of $r/\sigma$ for $\sigma >1$. We conclude that the error in
the pair potential can be written in the scaling form
\begin{equation}
\delta U_G(\rr,\rr', \sigma) = {1\over \sigma^5} V_5((\rr-\rr')/\sigma)
\end{equation}
for $\sigma >1$. $V_5$ {\sl does}\/ depend on the direction of the
relative displacement with respect to the lattice.  The error
increases strongly for $r/\sigma<2$, however smaller distances in our
simulations will be within the core of the Lennard-Jones potential and
will not be sampled. Due to the regularity of $V_5$ one could also
improve accuracy of the simulation by subtracting the error off of the
real space potential after parameterizing it with splines.  For
$\sigma<1$ aliasing errors are increasingly important, adding an
sinusoidal contribution to the error as expected from
eq.~(\ref{sinusoid}).  $\sigma=1$ generates errors in the potential
which are $O(10^{-4})$.

One simulates a system described by the energy eq.~(\ref{U2}) with the
constraint eq.~(\ref{constraint}) with the Metropolis method by
introducing two independent Monte Carlo updates: Plaquette updates,
which satisfy $D\delta E=0$, consist of the coupled update of the 4
links forming a plaquette \cite{vincent} of the cubic lattice. On each
of these links the field is modified by the same amount $\Delta$, so
that the flux of $E$ at each node remains constant. With the addition
of nearest neighbor interactions, eq.~(\ref{locale}), calculation of
the energy change requires the values of the field from 12 links.
Motion of a particle is possible if a local update of the field is
performed simultaneously such that $ D \delta E = \delta \rho$ where
$\delta \rho$ is the localized charge fluctuation.

We implemented a simulation of TIP3P water using Gaussian
interpolation due to the superior convergence properties at higher
accuracies. We work in units of the mesh size. Three dimensional
Gaussians, with $\sigma=1$, are calculated as direct products of one
dimensional Gaussians each truncated at $\lambda=4.3$. The three atoms
of each molecule are interpolated together. The small error,
$O(10^{-8})$, in charge conservation is corrected on the grid point
nearest the oxygen atom.  In this way we insure that charge is
conserved in the algorithm to machine precision.  We perform a trial
move and re-perform both interpolation and charge correction steps.
This gives us a localized charge fluctuation $\delta \rho$.  We
generate a local field modification in a box enclosing the original
and final sites: We use a $\delta E$ such that $\sum_i \delta E_i^2/2$
is minimum while respecting $D\delta E = \delta \rho$.  Outside of the
region where $\delta \rho=0$ we impose that $\delta E=0$.  This leads
to a {\it small}\/ Poisson problem (with zero flux boundary
conditions) within the interpolation volume which can be solved using
the FFTW library \cite{FFTW05}.  Lennard-Jones and ${\rm erfc}$
interactions were truncated at a distance of 9{\AA}.  We use Monte
Carlo updates in both the position and orientation of the molecules,
tuned to give an acceptance rate of about $40\%$. For each update of a
molecule we perform 100 plaquette updates.  Simulations were performed
at $300K$ at constant volume. Due to the simplicity of the plaquette
updates compared with the the calculation of the ${\rm erfc}$
interaction, they take a small part of the CPU time.

\begin{figure}[htb]
  \includegraphics[scale=.45,] {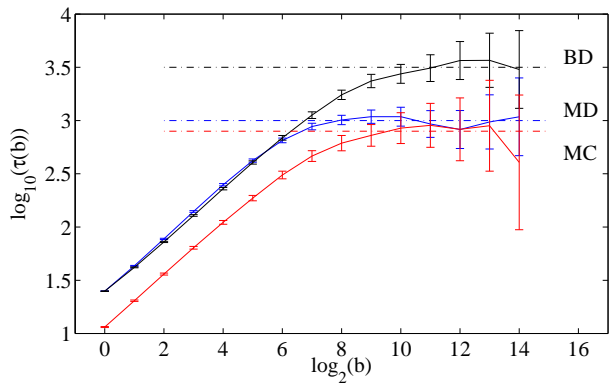}
  \caption{Blocking analysis of the energy with Monte Carlo (MC),
    molecular dynamics (MD) and Brownian dynamics (BD) to estimate the
    energy autocorrelation time.  The estimated $\tau$ for each method
    is indicated with a dot-dashed line.  }
  \label{fig:block}
\end{figure}

We compared our Monte Carlo simulation of TIP3P with a molecular
dynamics simulation \cite{gromacs} using using a Langevin thermostat,
friction 1 ps$^{-1}$, integration time step 1 fs.  We used a cubic box
of side 18.62 \AA \, and a grid of $20^3$ sites for the Monte Carlo.
We measured the autocorrelation time of the potential energy, $V$
using blocking \cite{blocking}, after removing the energy in the
transverse electric field in the Monte Carlo runs. A set of recordings
corresponding to $T$ sweeps or time steps is averaged in blocks of
$b=2^m$ recordings, to estimate the mean potential energy, $\langle V
\rangle $, and a running estimate in the error in $\langle V \rangle
$, $\tilde \sigma(b)$.  For large blocking factors $\tilde \sigma(b)$
saturates to a constant $\tilde \sigma_v$; the integrated
autocorrelation time is given by
\begin{math}
  \tau =  \tilde \sigma_v^2 T/2 \langle  V^2 -\langle V \rangle^2 \rangle
\end{math}. In order to obtain reasonable statistics for the dynamics 
we simulated a small system of $216$ particles for several thousand
$\tau$.  The running estimate of $\tau(b)$ is plotted in
Figure~\ref{fig:block}.  We estimate that $\tau=1100$ for molecular
dynamics and $\tau=800$ for Monte Carlo.  We also performed Brownian
dynamics simulations with the time step equal to $1/10$ of the
stability limit using an Euler integrator finding $\tau=3200$.

To conclude, we have implemented a Monte Carlo algorithm for the
simulation of TIP3P.  Each Monte Carlo time step implies two
interpolations per charge plus calculation of a localized current.
Each molecular dynamics step requires one charge interpolation and
then three extrapolations for the force plus solution of the Poisson
equation.  The total complexity of the interpolation steps is very
similar in molecular dynamics (particularly multigrid) codes and in
our Monte Carlo formulation.  The integrated autocorrelation time with
our algorithm is comparable to simulations performed using molecular
dynamics when measured in sweeps. CPU time comparisons were less
favorable to our code since the (Fourier based) Gromacs package
contains contains extensive assembly routines for interpolation which
we did not implement; the present Monte Carlo implementation is an
order of magnitude slower.

The authors in \cite{shaw} reduced the range of the interpolation step
by introducing a second on-lattice convolution after interpolating
charges to the grid.  Similar techniques are possible here if we
modify the kernel $K$ so as to include a extra spreading step; charge
interpolation then becomes cheaper. In our algorithm this
simplification in charge motion is balanced by an increase in the
complexity of plaquette updates.

Our algorithm has the important advantage over other codes of being
purely local, and thus easily implemented on parallel computers with
limited interprocessor communication, such as is the case on low cost
clusters. With a Monte Carlo algorithm there is also an enormous gain
in flexibility in heterogenous environments: In simulations of a
biomolecule or an interface most of the water molecules play the role
of distant spectator, even if they provide the majority of charge
centers. In Monte Carlo it is trivial to bias moves towards
interesting degrees of freedom, or even introduce cluster \cite{rick}
or multi-step \cite{berne} updates; with molecular dynamics
multi-scale and multi-step algorithms are difficult to implement and
prone to instability.

\bibliography{polar}

\end{document}